\documentclass[a4]{epl2}
\usepackage{graphicx}
\usepackage{epsfig}
\usepackage{color}
\usepackage{graphicx}
\usepackage{epstopdf}
\usepackage{amsmath}
\usepackage{float}

\usepackage{fancyhdr}
\title{\textcolor[rgb]{0.00,0.00,1.00}{Electrodynamics with quaternionic mass}}

\author{A. I. Arbab\inst{}\footnote{arbab.ibrahim@gmail.com}}
\date{today}
\institute{
Department of Physics, College of Science, Qassim University, P.O. Box 6644,  Buraidah 51452,  Saudi Arabia}


\abstract{New electrodynamics with quaternionic mass is found to yields interesting results.  The quaternionic mass involves longitudinal as well as transverse (vector) masses. Because of these two masses, an application of a magnetic field in a system induces electric charge and current densities in the system. Frank Wilczek axion electrodynamics is derived from this new electrodynamics bearing a quantum signature. A new version of Maxwell-Proca equations that are  invariant under the gauge transformations is derived from the new electrodynamics. New electric and magnetic fields that are different from the standard formalism are proposed to describe massive electrodynamics. Axion electrodynamics are shown to be equivalent to Proca electrodynamics. The energy and momentum conservation equations are not influenced by the quaternionic mass. }

\begin{document}
\date{today}
\maketitle
\baselineskip=2\baselineskip

\section{\textcolor[rgb]{0.00,0.00,1.00}{Introduction}}
Maxwell's theory had laid down the foundations of electromagnetic phenomena we now know \textcolor[rgb]{0.00,0.00,1.00}{\cite{griffth}}. Maxwell's equations were firstly expressed in quaternions in terms of 20 variables. By eliminating all scalars in the original Maxwell's equations, Heaviside and Gibbs reduced them to 4 equations which we now know. Further, Proca extended Maxwell's equations to include massive vector bosons \textcolor[rgb]{0.00,0.00,1.00}{\cite{proca}}. Quaternionic representation of a physical quantity generalizes the real representation to generalized complex representation. These include the particle electromagnetic field and electric current density which are the main constituents of Maxwell equations.

The modern understanding of particle interactions is made via the exchange of a mediator (gauge boson). The mass of the gauge boson is generated by employing the spontaneous symmetry breaking prescription (the Higgs mechanism). The electromagnetic gauge boson is the photon. The photon is treated as a massless boson to save the gauge invariance of Maxwell's theory. The photon has a dual nature if it behaves like a particle. Its electrodynamic equations in this instance should be governed by quantum mechanics in addition  to Maxwell equations. For such a description, we apply quantum rules for the massive photon, firstly treating the electromagnetic field as its wavefunction once, and secondly the electromagnetic potential as its wavefunction.

Using quaternions, one can extend Maxwell's equation to involve an additional scalar and a massive photon \textcolor[rgb]{0.00,0.00,1.00}{\cite{quantized}}.  The mass of the particle is normally represented by a real number. If we allow the photon's mass to be represented by a quaternion, additional interesting consequences will arise. The photon with mass requires quantum mechanics and electrodynamics to be combined in one framework. The new derived quantum electrodynamics  yields the axion electrodynamics proposed recently by Frank Wilczek \textcolor[rgb]{0.00,0.00,1.00}{\cite{wilczek,wilczek2}}. Maxwell-Proca equations that are invariant under the gauge transformations are  found to result from this formalism. Other new sets of modified Maxwell's equations are derived.

We would like in this work to explore these interesting consequences by allowing the photon mass to be a quaternion. The resulting equations are capable of explaining some of the experimental facts now observed in the field of condensed matter physics \textcolor[rgb]{0.00,0.00,1.00}{\cite{insulator}}. The application of a magnetic field is found to induce electric charge and current densities in the system.

This paper is organized as follows: we introduce in Section 2 the concept of quaternionic mass and work out the electrodynamics associated with it.  In Section 3, we apply the quaternionic momentum eigen-value equation (Dirac-type) to derive the quantum mechanics of a massive photon, where the electromagnetic field is considered to be its wavefunction. This leads to a system of equations similar to Maxwell's electrodynamics. A transformation is made to yield the corresponding electrodynamics that involves Maxwell-Proca equations and their generalization. Section 4 explores a Dirac-like quantum mechanical model of the massive photon by associating a quaternionic wavefunction consisting of scalar and vector components.  This model is found to generalize the axion and Proca electrodynamics. We introduce new electrodynamics of based on the momentum eigen-value equation  in Section 6 where the electromagnetic potential is taken as the photon's wavefunction. The derived equations are found to generalize Maxwell-Proca equations too. A Dirac-like quantum mechanics for massive photon employing the electromagnetic potential as its wavefunction is studied in Section 7. We end our paper with concluding remakes in Section 8.

\section{Electrodynamics with quaternionic mass}

We have studied earlier massive electrodynamics  using quaternion formalism \textcolor[rgb]{0.00,0.00,1.00}{\cite{quantized}}. In this study we treat the particle mass as a real quantity. Massive electrodynamics was also studied by Proca but his theory was not gauge invariant \textcolor[rgb]{0.00,0.00,1.00}{\cite{proca}}. However, our present formalism is gauge invariant. We would like now to consider massive electrodynamics where the mass is quaternionic. To this end, one writes\footnote{$\tilde{F}=\tilde{\nabla}\tilde{A}$\,, where $\tilde{A}=(\frac{i}{c}\varphi\,,\vec{A})$ and  $\tilde{\nabla}=(\frac{i}{c}\frac{\partial}{\partial t}\,,\vec{\nabla})$.}
\begin{equation}
\tilde{P}\tilde{F}=c\tilde{M}\tilde{F}\,,\,\,\,\, \tilde{P}=\left(\frac{i}{c}\,E\,,\vec{p}\right),\,\,\,\, \tilde{F}=\left(\Lambda\,, \vec{B}-\frac{i}{c}\,\vec{E}\right),\qquad \tilde{M}=(im_\ell\,, \vec{m}_t)\,,
\end{equation}
where  $\vec{E}$ and $\vec{B}$ are the electric and magnetic fields, and $\Lambda$ expresses the Lorenz gauge condition,
\begin{equation}
\vec{\nabla}\cdot\vec{A}+\frac{1}{c^2}\,\frac{\partial\varphi}{\partial t}=-\Lambda\,.
\end{equation}
The product of two quaternions, $\tilde{A}=(a_0\,,\vec{a})$ and $\tilde{B}=(b_0\,,\vec{b})$ is given by \textcolor[rgb]{0.00,0.00,1.00}{\cite{product}}
\begin{equation}\tag{A}
\tilde{A}\tilde{B}=(a_0b_0-\vec{a}\cdot\vec{b}\,,a_0\vec{b}+\vec{a}b_0+\vec{a}\times\vec{b})\,.
\end{equation}
Expanding eq.(1) using the above quaternionic product rule, we obtain
\begin{equation}
\vec{\nabla}\cdot\vec{E}=\frac{\partial\Lambda}{\partial t}-\frac{c^2}{\hbar}\vec{m}_t\cdot\vec{B}\,,
\end{equation}
\begin{equation}
\vec{\nabla}\cdot\vec{B} =\frac{1}{\hbar}\, \vec{m}_t\cdot\vec{E}+\frac{m_\ell c}{\hbar}\, \Lambda\,,
\end{equation}
\begin{equation}
\vec{\nabla}\times\vec{E}=-\frac{\partial\vec{B}}{\partial t}-\frac{c^2}{\hbar}\Lambda\vec{m}_t-\frac{m_\ell c}{\hbar}\,\vec{E}-\frac{c^2}{\hbar}\,\vec{m}_t\times\vec{B} \,,
\end{equation}
and
\begin{equation}
\vec{\nabla}\times\vec{B}=\frac{1}{c^2}\frac{\partial\vec{E}}{\partial t}-\vec{\nabla}\Lambda+\frac{\vec{m}_t}{\hbar}\,\times\vec{E}-\frac{m_\ell c}{\hbar}\,\vec{B}\,.
\end{equation}
The above system of equations is recently derived for real mass in the framework of coupling the quaternionic Dirac field with photons \textcolor[rgb]{0.00,0.00,1.00}{\cite{matter}}. A complementary electrodynamics is also derived in \textcolor[rgb]{0.00,0.00,1.00}{\cite{quantized}}. It is interesting to notice that eqs.(3) - (6) are gauge invariant. The above electrodynamics reduces to least modified electrodynamics recently derived electrodynamics when $\vec{m}_t=0$ and $m_\ell=0$ \textcolor[rgb]{0.00,0.00,1.00}{\cite{modified-maxwell}}. As evident from eqs.(3) and (6), one can define an electric charge density induced by the vector mass  as $\rho_a=-\frac{\vec{m}_t\cdot\vec{B}}{\mu_0\hbar}$\, and an electric current density that is perpendicular to the electric field by $\frac{\vec{m}}{\mu_0}\times\vec{E}$\,. The existence of these two quantities give rise to interesting results occurring in topological insulators \textcolor[rgb]{0.00,0.00,1.00}{\textcolor[rgb]{0.00,0.00,1.00}{\cite{insulator}}}. The above electrodynamics is invariant under duality transformations if $\Lambda=0$.

Notice that a massive particle has a vanishing transverse mass when it is completely free. The particle interacts with a momentum given by $\vec{p}=c\,\vec{m}_t$. Thus, when comparing this with the minimal coupling method for a charged particle interacting with an electromagnetic field, where, $\vec{p}\,'=\vec{p}+q\vec{A}$, then $c\,\vec{m}_t=q\vec{A}$. Therefore, the interaction of  the electromagnetic field with any other field defined by $\chi$ will be governed by eqs.(3) - (6) such that  $c\,\vec{m}=\hbar\vec{\nabla}\chi$\,.

Thus, eqs.(3) - (6) show that the masses are the source of the electromagnetic field. This helps us  understand why the neutron has electromagnetic properties although it is neutral. This is truly possible if one relates the transverse mass to the particle's spin. The above mass interaction of the electromagnetic fields allows gravity to be included too. However, in  Einstein's theory of gravitation, the influence  of gravity on a theory is to use the  covariant derivatives instead of the normal derivatives via the so-called the General Covariance Principle \textcolor[rgb]{0.00,0.00,1.00}{\cite{weinberg}}. This replacement brings about some extra terms in the theory that would represent the interaction of the initial fields with gravity.

The energy conservation equation associated with eqs.(3) - (6) is given by
\begin{equation}
\frac{\partial u}{\partial t}+\vec{\nabla}\cdot\vec{S}=0\,,\qquad u=\frac{B^2+\Lambda^2}{2\mu_0}+\frac{1}{2}\,\varepsilon_0E^2\,,\qquad \vec{S}=\mu_0^{-1}(\vec{E}\times\vec{B}+\Lambda\vec{E})\,,
\end{equation}
which is independent of the masses, $\vec{m}_t$ and $m_\ell$. Equation (7) reveals that there is no energy loss due to the system described by the quaternionic mass.
Therefore, these two masses are but some kinds of gauge fields.

The momentum conservation equation of the above electrodynamics system can be obtained from eqs.(3) - (6) which reads
\begin{equation}\tag{7a}
\frac{\partial}{\partial t}\,\left(\vec{E}\times\vec{B}-\Lambda\,\vec{E}\right)_i+\partial_j\left(\frac{\delta_{ij}}{2}(E^2+c^2B^2-c^2\Lambda^2)+\varepsilon_{ijk}c^2(\Lambda\,B_k)-(E_iE_j+c^2B_iB_j)\right)=0\,,
\end{equation}
which can be expressed as
\begin{equation}\tag{7b}
-\frac{\partial g_i}{\partial t}+\partial_j\,\sigma_{ij}=0\,,
\end{equation}
where
\begin{equation}\tag{7c}
\begin{split}
\sigma_{ij}=\varepsilon_0(E_iE_j+c^2B_iB_j)-\frac{\delta_{ij}}{2}\,\varepsilon_0(E^2+c^2B^2-c^2\Lambda^2)-\varepsilon_{ijk}\,\mu_0^{-1}\,\Lambda\,B_k)\,, \\
g_i=\varepsilon_0\left(\vec{E}\times\vec{B}-\Lambda\,\vec{E}\right)_i\,.
\end{split}
\end{equation}
Here $g_i$ and $\sigma_{ij}$ are the momentum density and stress tensor of the electromagnetic field involving the effect of the scalar $\Lambda$\,. It is now apparent that the scalar field $\Lambda$ increases the energy density of the electromagnetic field at the expense of losing momentum density, and vice versa.
Once again the mass of the particle does not influence the momentum conservation equation. The system appears as if it has zero mass. It is pertinent to mention that the energy gained by $\Lambda$, eq.(7), is lost by the momentum, eq.(7a). The stress tensor in eq.(7c) is no longer symmetric. A method to obtain a symmetric stress is provided by Belinfante \textcolor[rgb]{0.00,0.00,1.00}{\cite{nonsymm}}. Recall that the energy-momentum tensor obtained from the electromagnetic Lagrangian density was not symmetric too. The above energy and momentum conservation equations agree with our electrodynamics with vanishing masses ($\vec{m}_t=0$ and $m_\ell=0$) \textcolor[rgb]{0.00,0.00,1.00}{\cite{modified-maxwell}}.
A manipulation with eqs.(3) - (6) reveals that the scalar $\Lambda$ satisfies the Klein-Gordon equation,
\begin{equation}
\frac{1}{c^2}\frac{\partial^2\Lambda}{\partial t^2}-\nabla^2\Lambda+\left(\frac{Mc}{\hbar}\right)^2\Lambda=0\,,
\end{equation}
where
\begin{equation}
M^2=m_t^2-m_\ell^2\,.
\end{equation}
Remarkably, one can define the mass $M$ as a rest-mass of the field. Therefore, eqs.(3) - (6) reflect the duality of the massive photon where the electromagnetic field of the massive photon shows matter as well as wave aspects. The two masses should be related to the wave and particle natures of the massive photon. Interestingly, a massless particle (field) has $|\vec{m}_t|=m_\ell$\,, and not necessarily vanishing two masses. This will clearly show that mass has another genuine meaning.

\subsection{Axion electrodynamics and quantum mechanics}

The axion  is a particle proposed to  resolve the strong CP problem in quantum chromodynamics (QCD), but lately is thought to contribute to the dark matter in the universe. Wilczek found the electrodynamics describing the evolution of axions by the modified Maxwell's equations \textcolor[rgb]{0.00,0.00,1.00}{\cite{wilczek,wilczek2}}
\begin{equation}
\vec{\nabla}\cdot\vec{E}=-\kappa\vec{\nabla}a\cdot\vec{B}\,,
\qquad\qquad
\vec{\nabla}\times\vec{E}=-\frac{\partial\vec{B}}{\partial t} \,,
\end{equation}
\begin{equation}
\vec{\nabla}\cdot\vec{B} =0\,,
\qquad\qquad
\vec{\nabla}\times\vec{B}=\frac{1}{c^2}\frac{\partial\vec{E}}{\partial t}+\kappa\left(\frac{\partial a}{\partial t}\,\vec{B}+\vec{\nabla} a\,\times\vec{E}\right)\,,
\end{equation}
where $a$ and $\kappa$ are the axion field, and the  axion–photon coupling, respectively.
It is interesting to see that eqs.(3) - (6) are analogous to the axion-electrodynamics describing the axion-photon interaction, if one sets \textcolor[rgb]{0.00,0.00,1.00}{\cite{wilczek,axion}}
\begin{equation}
\vec{m}_t=\hbar g_{a\gamma}\,\vec{\nabla}a\,,\qquad m_\ell c=-\hbar g_{a\gamma}\frac{\partial a}{\partial t}\,\,\qquad\Rightarrow\qquad \vec{m}_t=-m_\ell c\left(\frac{\vec{\nabla}a}{\dot a}\right)\,,\qquad \dot a=\frac{\partial a}{\partial t}\,,
\end{equation}
 Notice that  \textcolor[rgb]{0.00,0.00,1.00}{\cite{axion}} uses $g_{a\gamma}a=\frac{\kappa}{c}\,\theta$ instead. Hence, eqs.(3) - (6) can be seen as the quantized axion electrodynamics. They also generalize the axion electrodynamics to allow magnetic current and charge densities.
One can compare the axion electrodynamics studied by Tiwari adopting a duality  perspective with our present formalism  \textcolor[rgb]{0.00,0.00,1.00}{\cite{tiwari}}. His equations, eqs.[39] - [42], are in agreement with our equations, eqs.(3) - (6),  upon substituting $g\vec{W}\rightarrow \frac{\vec{m}_t}{\hbar}$ and $gW_0\rightarrow \frac{m_\ell c}{\hbar}$, $\vec{W}$ and $W_0$ are the duality gauge four-vector, and $g$ is the coupling constant of this field with the electromagnetic field.

One can consider the interaction of the quantum particle (electron) with the axion field by substituting eq.(12) in the quaternionic momentum eigen-value equation with a quaternionic mass defined by $\tilde{P}\tilde{\Psi}^*=c\tilde{M}\tilde{\Psi}^*$  \textcolor[rgb]{0.00,0.00,1.00}{\cite{qm_mass}}. This substitution yields
\begin{equation}
\frac{1}{c^2}\,\frac{\partial\psi_0}{\partial t}+\vec{\nabla}\cdot\vec{\psi}=0\,,
\end{equation}
\begin{equation}
  \frac{\partial\vec{\psi}}{\partial t}+\vec{\nabla}\psi_0+c^2g_{ap}\vec{\nabla}a\times\vec{\psi}=0\,\,,
\end{equation}
\begin{equation}
\vec{\nabla}\times\vec{\psi}-g_{ap}\,\left(\frac{\partial a}{\partial t}\,\vec{\psi}+(\vec{\nabla}a)\,\psi_0\right)=0\,,
\end{equation}
and
\begin{equation}
\left(\frac{\partial a}{\partial t}\right)\,\psi_0+c^2(\vec{\nabla}a)\cdot\vec{\psi}=0\,,
\end{equation}
where $g_{ap}$ describes the axion-particle (electron) coupling. Equation (12)  shows that the axion field imparted momentum and energy  to the particle (electron) given,  respectively by,
\begin{equation}
c\,\vec{m}_t=c\hbar g_{ap}\vec{\nabla}a\,,\qquad\qquad\qquad m_\ell c^2=-\hbar cg_{ap}\frac{\partial a}{\partial t}\,,
\end{equation}
where $\vec{m}_t$ is a constant vector.
Remarkably,  the interaction of the axion field with a quantum particle is made by the minimal coupling prescription by setting $q\vec{A}\rightarrow c\vec{m}_t$ and $q\varphi\rightarrow m_\ell c^2$\,.

\section{Quaternionic Maxwell equation}

Let us now assume the massive electromagnetic field, $\tilde{F}$ satisfies  the Dirac's momentum eigen-value equation. We further let  the massive photon to have  a quaternionic mass.  To this end one writes
\begin{equation}
\tilde{P}\,\tilde{\gamma}\,\tilde{F}=c\tilde{M}\tilde{F}\,,\qquad \tilde{\gamma}=(i\beta\,,\vec{\gamma})\,.
\end{equation}
Expanding eq.(18) and equating the real and imaginary parts in the two sides of the resulting equations, yield
\begin{equation}
\vec{\nabla}\cdot(-c\vec{\gamma}\Lambda+\beta\vec{E}+c\vec{\gamma}\times\vec{B})=-\frac{\partial}{\partial t}(\beta\Lambda-\frac{\vec{\gamma}}{c}\cdot\vec{E})-\frac{m_\ell c^2}{\hbar}\,\Lambda+\frac{c}{\hbar}\vec{m}_t\cdot\vec{E}\,,
\end{equation}
\begin{equation}
\vec{\nabla}\cdot\,(c\beta\vec{B}-\vec{\gamma}\times\vec{E})=\frac{\partial}{\partial t}(\vec{\gamma}\cdot\vec{B})-\frac{c^2}{\hbar}\,\vec{m}_t\cdot\vec{B}\,,
\end{equation}
\begin{equation}
-\vec{\nabla}\times(-c\vec{\gamma}\Lambda+\beta\vec{E}+c\vec{\gamma}\times\vec{B})+\vec{\nabla}(c\vec{\gamma}\cdot\vec{B})-\frac{1}{c}\,\frac{\partial}{\partial t}(c\beta\vec{B}-\vec{\gamma}\times\vec{E})=-\frac{c}{\hbar}\,\vec{m}_t\times\vec{E}+\frac{m_\ell c^2}{\hbar}\,\vec{B}\,,
\end{equation}
and
\begin{equation}
-\vec{\nabla}\times(c\beta\vec{B}-\vec{\gamma}\times\vec{E})+\vec{\nabla}(c\beta\Lambda-\vec{\gamma}\cdot\vec{E})+\frac{1}{c}\,\frac{\partial}{\partial t}(-c\vec{\gamma}\,\Lambda+\beta\vec{E}+c\vec{\gamma}\times\vec{B})=\frac{c^2}{\hbar}\,\vec{m}_t\Lambda-\frac{m_\ell c}{\hbar}\,\vec{E}-\frac{c^2}{\hbar}\, \vec{m}_t\times\vec{B}\,.
\end{equation}
The above equations represent the electric field of the photon having wave and particle natures. To this description one defines
\begin{equation}
\vec{E}_M=\vec{E}+c\beta\vec{\gamma}\times\vec{B}-c\beta\vec{\gamma}\Lambda\,,\qquad\qquad
\vec{B}_M=\vec{B}-\frac{\beta\vec{\gamma}}{c}\times\vec{E}\,,\qquad \qquad\Lambda_M=\Lambda-\frac{\beta\vec{\gamma}}{c}\cdot\vec{E}\,,
\end{equation}
as the effective electric and magnetic fields, and the scalar field,
so that eqs.(19) - (22) become
\begin{equation}
\vec{\nabla}\cdot\vec{E}_M=-\frac{\partial\Lambda_M}{\partial t}-\mu\,c\,\Lambda_M\,,
\end{equation}
\begin{equation}
\vec{\nabla}\cdot\,\vec{B}_M=-\frac{1}{c}\frac{\partial}{\partial t}(\beta\vec{\gamma}\cdot\vec{B}_M)-\beta\mu\,\vec{\gamma}\cdot\vec{B}_M\,,
\end{equation}
\begin{equation}
\vec{\nabla}\times\vec{E}_M=-\frac{\partial\vec{B}_M}{\partial t}+\vec{\nabla}(c\beta\vec{\gamma}\cdot\vec{B}_M)-\beta\,c\vec{B}_M\,,
\end{equation}
and
\begin{equation}
\vec{\nabla}\times\vec{B}_M=\frac{1}{c^2}\,\frac{\partial\vec{E}_M}{\partial t}+\vec{\nabla}\Lambda_M+\frac{\beta\mu }{c}\,\vec{E}_M\,,
\end{equation}
where
\begin{equation}
\vec{\gamma}\,m_\ell=\beta\,\vec{m}_t\,.
\end{equation}
Recall that in Dirac's theory, one has $\vec{v}=c\,\beta\,\vec{\gamma}$ and the eigen-value of $\beta$ are $\pm\,1$, so that eqs.(23) becomes
\begin{equation}
\begin{split}
\vec{E}_M=\vec{E}+\vec{v}\times\vec{B}-\vec{v}\,\Lambda\,,\qquad\qquad
\vec{B}_M=\vec{B}-\frac{\vec{v}}{c^2}\times\vec{E}\,,\\ \Lambda_M=\Lambda-\frac{\vec{v}}{c^2}\cdot\vec{E}\,,\qquad\qquad \vec{v}\cdot\vec{B}=0\,.
\end{split}
\end{equation}
Interestingly, eq.(28) now reads
\begin{equation}\tag{28a}
m_\ell\vec{v}=c\,\vec{m}_t\,,
\end{equation}
that reveals the momentum of the particle is equal to that of a wave. This indicates that $\vec{m}_t$ represent a wave's mass whereas $m_\ell$ a particle's mass. This is a mathematical realization of the wave-particle duality principle suggested by de Broglie. The relation in eq.(28a) was already found when a quantum mechanical aspect of a particle is considered \textcolor[rgb]{0.00,0.00,1.00}{\cite{arbab et al}}.

If we now apply eq.(29) in eqs.(24) - (27), one finds
\begin{equation}
\vec{\nabla}\cdot\vec{E}_M=-\frac{\partial\Lambda_M}{\partial t}\mp\mu\,c\,\Lambda_M\,,
\end{equation}
\begin{equation}
\vec{\nabla}\cdot\,\vec{B}_M=0\,,
\end{equation}
\begin{equation}
\vec{\nabla}\times\vec{E}_M=-\frac{\partial\vec{B}_M}{\partial t}\mp\mu\,c\,\vec{B}_M\,,
\end{equation}
and
\begin{equation}
\vec{\nabla}\times\vec{B}_M=\frac{1}{c^2}\,\frac{\partial\vec{E}_M}{\partial t}+\vec{\nabla}\Lambda_M\pm\frac{\mu }{c}\,\vec{E}_M\,,\qquad \mu=\frac{m_\ell c}{\hbar}\,.
\end{equation}
The above equations are but the massive field Maxwell's equation in a moving frame. They are the quantized Maxwell's equations that were recently derived \textcolor[rgb]{0.00,0.00,1.00}{\cite{quantized}}. The electromagnetic field in the above equations has four components describing the particle and antiparticle states.

\section{Maxwellian quantum mechanics with quaternionic mass}

Let us now consider a Dirac-like quaternionic momentum eigen-value equation, and consider the mass to be quaternionic. A real mass equation is recently considered and shown to lead to Maxwell-like equations of quantum mechanics  \textcolor[rgb]{0.00,0.00,1.00}{\cite{quantized}}. The Dirac-like quaternionic momentum eigen-value equation with quaternionic mass is expressed as
\begin{equation}
\tilde{P}\,\tilde{\gamma}\,\tilde{\Psi}=c\tilde{M}\,\tilde{\Psi}\,,\qquad\qquad \tilde{\gamma}=(i\beta\,, \vec{\gamma}),\qquad \qquad \tilde{\Psi}=(\frac{i}{c}\,\psi_0\,,\vec{\psi})\,,
\end{equation}
where $\vec{\gamma}$ and $\beta$ are the Dirac matrices \textcolor[rgb]{0.00,0.00,1.00}{\cite{bjorken}}. The quaternionic wavefunction $\tilde{\Psi}$ represents the state of the particle. It has four components that is analogous to Dirac's spinors. It is more akin to the scalar and vector representation of the electromagnetic field.

Expanding eq.(34) and equating the real and imaginary parts in the two sides of the  resulting equations to each other, yield
\begin{equation}
\vec{\nabla}\cdot\vec{E}_d=-\frac{\partial\Lambda_d}{\partial t}-\mu\,\psi_0-\frac{c^2}{\hbar}\,\vec{m}_t\cdot\vec{\psi}\,,
\end{equation}
\begin{equation}
\vec{\nabla}\cdot\vec{B}_d =0\,,
\end{equation}
\begin{equation}
\vec{\nabla}\times\vec{E}_d=-\frac{\partial\vec{B}_d}{\partial t}-\frac{c^2}{\hbar}\, \vec{m}_t\times\vec{\psi}\,,
\end{equation}
\begin{equation}
\vec{\nabla}\times\vec{B}_d=\frac{1}{c^2}\frac{\partial\vec{E}_d}{\partial t}+\vec{\nabla}\Lambda_d-\mu\,\vec{\psi}-\frac{\vec{m}_t}{\hbar}\,\psi_0\,,
\end{equation}
where
\begin{equation}
\vec{E}_d=-(c\beta\vec{\psi}+\vec{\gamma}\,\psi_0)\,,\qquad \vec{B}_d=\vec{\gamma}\times\vec{\psi}\,,\qquad \Lambda_d=\frac{\beta}{c}\,\psi_0+\vec{\gamma}\cdot\vec{\psi}\,.
\end{equation}
Equation (35) - (39) show a deep relationship between Dirac and Maxwell fields. Equation (39) shows how the matter and electromagnetic fields are related. Notice that $\vec{E}_d\cdot\vec{B}_d=0$. In electrodynamics, the electric and magnetic fields are defined by the derivatives of the scalar and vector potentials. Here the matter \emph{electroinertial} ($\vec{E}_d$) and \emph{magnetoinertial} ($\vec{B}_d$) fields are constructed directly from the particle wavefunctions. It is interesting to see that for $\vec{m}_t=$, $m_\ell=0$ and $\Lambda=0$, we obtain Maxwell - Proca - like equations with $\vec{\psi}\rightarrow\vec{A}$ and $\psi_0\rightarrow\varphi$\,.

The right-hand sides of eqs.(35) - (39) express the interaction of the quantum particle fields with the external field via $\vec{m}$. They can also be compared with eqs.(3) - (6) of the Maxwell fields. Interestingly, eq.(35) - (38) were recently derived but with different definitions of the electric and magnetic fields \textcolor[rgb]{0.00,0.00,1.00}{\cite{maxwellian}}.

In quantum electrodynamics, the electric current and charged density in Maxwell's equations are defined by the Dirac's spinor (wavefunction) of the electron.  It is thus clear that particle mass can give rise to electromagnetic interactions. It is interesting to see $\vec{E}_d$ and $\vec{B}_d$ are invariant under the transformation
\begin{equation}
\vec{\psi}\,'=\vec{\psi}\pm\frac{\gamma}{c}\,\psi_0\,,\qquad\qquad \psi_0'=\psi_0\pm\beta\,\psi_0\,.
\end{equation}
Equations (35) - (38) are invariant under the above transformation if $\Lambda=0$, and
\begin{equation}
m_\ell \,\vec{\gamma}=-\vec{m}_t\,\beta\,.
\end{equation}
Since $\vec{\gamma}=\beta\vec{\alpha}$, then eq.(41) yields
\begin{equation}\tag{41a}
m_\ell \,\vec{\alpha}=\vec{m}_tI\,
\end{equation}
where in Dirac's theory, $\vec{v}=c\vec{\alpha}$ \textcolor[rgb]{0.00,0.00,1.00}{\cite{bjorken}} and $I$ is an identity matrix. Therefore, the longitudinal and transverse masses can enter Dirac equation via the above relation. Now eq.(41a) can be expressed as
\begin{equation}\tag{41b}
m_\ell \,\vec{v}=\vec{m}_tc\,.
\end{equation}
Interestingly, eq.(41b) expresses the equality of matter-wave momentum and particle momentum as demonstrated by de Broglie's hypothesis. In this case, $\vec{m}_t$ represents the wave vector mass and $m_\ell$ is the particle mass.

The corresponding Maxwell's equations analogous to eqs.(35) - (38) are obtained by the substitution, $\vec{E}_d\rightarrow\vec{E}$, $\vec{B}_d\rightarrow\vec{B}$, $\Lambda_d\rightarrow\Lambda$, $\psi_0\rightarrow \frac{m_\ell c}{\hbar}\,\varphi$, and  $\vec{\psi}\rightarrow \frac{m_\ell c}{\hbar}\,\vec{A}$. This substitution yields
\begin{equation}
\vec{\nabla}\cdot\vec{E}=-\frac{\partial\Lambda}{\partial t}-\mu^2\varphi-\mu\,c^2\frac{\vec{m}_t }{\hbar}\,\cdot\vec{A}\,,
\end{equation}
\begin{equation}
\vec{\nabla}\cdot\vec{B} =0\,,
\end{equation}
\begin{equation}
\vec{\nabla}\times\vec{E}=-\frac{\partial\vec{B}}{\partial t}-\mu\,c^2\frac{\vec{m}_t }{\hbar}\times\vec{A}\,,
\end{equation}
\begin{equation}
\vec{\nabla}\times\vec{B}=\frac{1}{c^2}\frac{\partial\vec{E}}{\partial t}+\vec{\nabla}\Lambda-\mu^2\vec{A}-\mu\frac{\,\vec{m}_t }{\hbar}\,\varphi\,.
\end{equation}
Equations (42) - (45) reduce to the Maxwell-Proca equations if we set, $\Lambda=0$ and $\vec{m}_t=0$ \textcolor[rgb]{0.00,0.00,1.00}{\cite{proca}}. It is shown by Tajmar that Proca electrodynamics is appropriate to account for the Meissner effect in superconductivity \textcolor[rgb]{0.00,0.00,1.00}{\cite{tajmar}}. Equations (42) - (45) satisfy the following equation
\begin{equation}\tag{B}
\vec{\nabla}\times\vec{J}_m+\frac{\partial}{\partial t}\,(\mu_0\vec{J}_e)=0\,,\qquad \vec{J}_e=\frac{\vec{m}_t}{\mu_0\hbar}\times\vec{E}+\frac{\mu}{\mu_0}\,\vec{B}\,,\qquad\vec{J}_m=\mu\,\vec{E}-\frac{c^2}{\hbar}\Lambda\vec{m}_t-\frac{c^2}{\hbar}\,\vec{m}_t\times\vec{B}\,,
\end{equation}
where $\vec{J}_e$ and $\vec{J}_m$ are the electric and magnetic currents associated with the above system. It is apparent from eqs.(42) - (45) that electric charge, electric and magnetic currents are induced. These are given by
\begin{equation}\tag{45a}
\rho^e_A=-\frac{\mu}{\mu_0\hbar}\,\vec{m}_t\cdot\vec{A}\,,\qquad\vec{J}^m_A=\mu\,c^2\frac{\vec{m}_t }{\hbar}\times\vec{A}\,,\qquad \vec{J}^e_A=-\mu\frac{\,\vec{m}_t }{\mu_0\hbar}\,\varphi\,.
\end{equation}
The above electrodynamics, eqs.(42) - (45), is  invariant under the gauge transformations
\begin{equation}\tag{a}
\vec{A}\,'=\vec{A}-\vec{\nabla}f\,,\qquad\qquad \varphi\,'=\varphi+\frac{\partial f}{\partial t}\,,
\end{equation}
provided that
\begin{equation}\tag{b}
\vec{m}_t\times\vec{\nabla}f=0\,,\qquad m_\ell c\vec{\nabla}f-\vec{m}_t\,\frac{\partial f}{\partial t}=0\,,\qquad c\vec{m}_t\cdot\vec{\nabla}f-m_\ell \frac{\partial f}{\partial t}=0\,.
\end{equation}
This is satisfied if the scalar function, $f$, satisfies the wave equation
\begin{equation}\tag{c}
\frac{1}{c^2}\frac{\partial^2 f}{\partial t^2}-\nabla^2f=0\,.
\end{equation}
Mathematical consistency of eq.(b) requires that $|\vec{m}_t|^2=m_\ell^2$\,.
Inasmuch as  the transverse mass $\vec{m}_t$ expresses the interaction of the particle with the external field, the  Proca - Maxwell equations with quaternionic mass should read, with $\Lambda=0$\,,
 \begin{equation}
\vec{\nabla}\cdot\vec{E}=-\mu^2\varphi-\mu\,\frac{c^2\vec{m}_t}{\hbar}\,\cdot\vec{A}\,,
\end{equation}
\begin{equation}
\vec{\nabla}\cdot\vec{B} =0\,,
\end{equation}
\begin{equation}
\vec{\nabla}\times\vec{E}=-\frac{\partial\vec{B}}{\partial t}-\mu\,\frac{c^2\vec{m}_t}{\hbar}\, \times\vec{A}\,,
\end{equation}
\begin{equation}
\vec{\nabla}\times\vec{B}=\frac{1}{c^2}\frac{\partial\vec{E}}{\partial t}-\mu^2\vec{A}-\mu\frac{\,\vec{m}_t}{\hbar}\,\varphi\,.
\end{equation}
Here the vector boson field is described by its longitudinal and transverse masses. Furthermore, the mass scale in Proca - Maxwell  theory is $\mu=\frac{m_\ell c}{\hbar}$\,\textcolor[rgb]{0.00,0.00,1.00}{\cite{proca}}. These suggested equations are new to the best of my knowledge. Equation (46) - (49) reduce to the standard  Maxwell's equation in vacuum when $m_\ell=0$ and $\vec{m}_t=0$\,. However, if we let $c\vec{m}_t\rightarrow q\vec{A}$ and $m_\ell c^2\rightarrow q\varphi$, we will obtain a nonlinear electrodynamics. These are
\begin{equation}
\vec{\nabla}\cdot\vec{E}=-\frac{q^2\varphi^3}{\hbar^2c^2}-\frac{q^2}{\hbar^2}\,\varphi A^2\,,
\end{equation}
\begin{equation}
\vec{\nabla}\cdot\vec{B} =0\,,
\end{equation}
\begin{equation}
\vec{\nabla}\times\vec{E}=-\frac{\partial\vec{B}}{\partial t}\,,
\end{equation}
\begin{equation}
\vec{\nabla}\times\vec{B}=\frac{1}{c^2}\frac{\partial\vec{E}}{\partial t}-\frac{2q^2\varphi^2}{\hbar^2c^2}\,\vec{A}\,.
\end{equation}
These will represent the coupling of the electromagnetic field with massive photons. Equations (50) and (53) suggest  non-linear electric gauge charge and current densities given by
\begin{equation}\tag{i}
\rho_g=-\frac{\varepsilon_0q^2\varphi^3}{\hbar^2c^2}-\frac{\varepsilon_0q^2}{\hbar^2}\,\varphi A^2\,,\qquad\qquad \vec{J}_g=-\frac{2\varepsilon_0q^2}{\hbar^2}\, \varphi^2\vec{A}\,.
\end{equation}
Applying the Lorenz gauge condition in eqs.(50) and (53) one finds
\begin{equation}\tag{ii}
\frac{1}{c^2}\frac{\partial^2\varphi}{\partial t^2}-\nabla^2\varphi+ac^2\varphi\,A^2+a\varphi^3=0\,,\,\, \frac{1}{c^2}\frac{\partial^2\vec{A}}{\partial t^2}-\nabla^2\vec{A}+2a\varphi^2\vec{A}=0\,,\qquad a=\frac{q^2}{\hbar^2c^2}\,.
\end{equation}
Recall that the theory of superconductivity was explained by employing the non-linear Schrodinger's equation with  linear London's equations of superconductivity. It would be interesting to explore the application of the above non-linear massive electrodynamics with the non-linear Schrodinger's equation. We remark that the above  system of equations will very soon find its application in the field of condensed matter and the allied sciences.

\section{Massive photon with quaternionic mass}

It is recently shown that the quaternionic momentum eigen - value equation with the  quaternion photon's field, $\tilde{A}$, led to similar equations as that  due to Maxwell quaternionic equation \textcolor[rgb]{0.00,0.00,1.00}{\cite{massive-photon}}. Let us study this theory with quaternionic mass by writing
\begin{equation}
\tilde{P}^*\tilde{A}=c\tilde{M}\tilde{A}\,,\qquad\qquad \tilde{A}=\left(\frac{i}{c}\,\varphi\,,\vec{A}\right),
\end{equation}
where $\tilde{M}$ is the photon quaternionic mass. We assume here the photon to be massive. It has longitudinal and transverse masses.

Expanding eq.(54) and equate the real and imaginary parts in the two sides of the resulting equations to each other yield
\begin{equation}
\vec{\nabla}\cdot\vec{A}+\frac{1}{c^2}\,\frac{\partial\varphi}{\partial t}=0\,,  \qquad\qquad c\vec{m}_t\cdot\vec{A}+m_\ell \varphi=0\,,
\end{equation}
and
\begin{equation}
\vec{E}=\frac{c^2\vec{m}_t}{\hbar}\,\times\vec{A}\,, \qquad\qquad \vec{B}=\frac{\vec{m}_t}{\hbar}\,\varphi+\frac{m_\ell c}{\hbar}\,\vec{A}\,.
\end{equation}
It is known that the electric and magnetic fields of a massless photon are zero. But eq.(56) reveals that the photon has neither electric field nor magnetic fields not because  it has no charge, but because it has no mass.

Manipulating eqs.(55) and (56), one finds the following equations
\begin{equation}
\vec{\nabla}\cdot\vec{E}=-\frac{c^2\vec{m}_t}{\hbar}\cdot\vec{B}\,,\qquad\qquad \vec{\nabla}\cdot\vec{B}=0\,,
\end{equation}
\begin{equation}
\vec{\nabla}\times\vec{E}=-\frac{\partial\vec{B}}{\partial t}\,,
\end{equation}
and
\begin{equation}
\vec{\nabla}\times\vec{B}=\frac{1}{c^2}\,\frac{\partial\vec{E}}{\partial t}+\frac{\vec{m}_t}{\hbar}\times\vec{E}+\frac{m_\ell c}{\hbar}\,\vec{B}\,,
\end{equation}
Equations (57) - (59) can be compared with eqs.(3) - (6) with $\Lambda=0$. They are similar to eqs.(10) and (11) of Wilczek axion electrodynamics \textcolor[rgb]{0.00,0.00,1.00}{\cite{wilczek,wilczek2}}.
It is thus very interesting that the transverse mass, $\vec{m}_t$ couples to the electric and magnetic fields giving rise to electric  charge density besides an electric current density. The electric current density, $\vec{J}_e=\frac{\vec{m}_t}{\mu_0\hbar}\times\vec{E}+\frac{m_\ell c}{\mu_0\hbar}\,\vec{B}$,  is non-dissipative.

The above electrodynamics is found to be obtained by considering an axion field interacting with photons.  Axion electrodynamics is found to be a promising description for topological insulators \textcolor[rgb]{0.00,0.00,1.00}{\cite{insulator}}. Recall that the CP violation in QCD was explained as due to the presence of the axion field. Owing to, eqs.(57) - (59), such a violation is a consequence of the photon wave-particle nature, particularly, when the transverse mass (dynamical mass) of the photon does not vanish. Note that eqs.(57) - (59) are not invariant under CP because of the fact that $m_\ell\ne0$. Hence, one may attribute the non-invariance  of CP transformation  to the photon being massive. Eqs.(57) - (59) violate the Time - Reversal symmetry too.

Interestingly, eqs.(57) - (59) are invariant under the gauge transformations. The electric and magnetic fields in eqs.(57) - (59) satisfy the Klein-Gordon equation
\begin{equation}\tag{56a}
\frac{1}{c^2}\frac{\partial^2\vec{E}}{\partial t^2}-\nabla^2\vec{E}+\left(\frac{Mc}{\hbar}\right)^2\vec{E}=0\,,\qquad\qquad
\frac{1}{c^2}\frac{\partial^2\vec{B}}{\partial t^2}-\nabla^2\vec{B}+\left(\frac{Mc}{\hbar}\right)^2\vec{B}=0\,.
\end{equation}
Equations (57) - (59), with the aid of eqs.(55) and (56), yield
\begin{equation}\tag{56b}
\vec{\nabla}\cdot\vec{E}=-\left(\frac{Mc}{\hbar}\right)^2\varphi\,,\qquad\qquad \vec{\nabla}\cdot\vec{B}=0\,,
\end{equation}
\begin{equation}\tag{56c}
\vec{\nabla}\times\vec{E}=-\frac{\partial\vec{B}}{\partial t}\,,\qquad\qquad
\vec{\nabla}\times\vec{B}=\frac{1}{c^2}\,\frac{\partial\vec{E}}{\partial t}-\left(\frac{Mc}{\hbar}\right)^2\vec{A}\,.
\end{equation}
Interestingly, these equations are but the Maxwell-Proca equations of a massive boson whose mass is $M$. These new Maxwell-Proca equations are but invariant under the gauge transformations. Remarkably, the axion electrodynamics is equivalent to the new Maxwell-Proca electrodynamics provided one sets $\vec{m}\propto\vec{\nabla}a$ and $m_\ell\propto\partial a/\partial t\,.$

We remark here that in the standard model, the masses of the gauge fields are set to zero because the presence of the gauge's mass spoils the gauge symmetry. However, with the new formulation, such an impasse can be lifted and a theory with massive fields can be proposed.

The electric and magnetic fields in eq.(56) are invariant under gauge transformation, eq.(a),  provided
\begin{equation}\tag{C}
\vec{\nabla}(m_\ell cf)-\frac{\partial }{\partial t}(\vec{m}_tf)=0\,,\qquad \vec{\nabla}\cdot(\vec{m}_tf) -\frac{1}{c^2}\frac{\partial }{\partial t}(m_\ell cf)=0\,,\qquad \vec{\nabla}\times(\vec{m}_tf)=0\,.
\end{equation}
The above system of equations is satisfied if the function $f$ satisfies the wave equation
\begin{equation}\tag{D}
\frac{1}{c^2}\frac{\partial^2 f}{\partial t^2}-\nabla^2f=0\,.
\end{equation}
Intriguingly, eq.(C) suggests that the gauge transformations can be viewed as a translation of the scalar and vector potentials to read
\begin{equation}\tag{E}
\vec{A}\,'=\vec{A}-\frac{c}{\hbar}\,\vec{m}_tf\,,\qquad\qquad \varphi\,'=\varphi+\frac{m_\ell c^2}{\hbar}f\,.
\end{equation}
Remarkably, there are no electric and magnetic fields associated with the additional scalar and vector potentials in eq.(E).

\section{New electromagnetic fields}

We now employ the Dirac's momentum eigen-value equation for a massive photon described by the quaternionic wavefunction, $\tilde{A}$. Let us further treat the mass as a quaternion quantity. This then reads
\begin{equation}
\tilde{P}^*\tilde{\gamma}\,\tilde{A}^*=c\tilde{M}\tilde{A}^*\,.
\end{equation}
Expand eq.(60) and equate the real and imaginary parts in the two sides of the resulting equations to each other, yield
\begin{equation}
\vec{\nabla}\cdot(-c\beta\vec{A}+\vec{\gamma}\,\varphi)+\frac{\partial}{\partial t}\left(\beta\frac{\varphi}{c}-\vec{\gamma}\cdot\vec{A}\right)=-\frac{m_\ell c}{\hbar}\,\varphi+\frac{c^2}{\hbar}\vec{m}_t\cdot\vec{A}\,,
\end{equation}
\begin{equation}
\vec{\nabla}\cdot\,(\vec{\gamma}\times\vec{A})=0\,,
\end{equation}
\begin{equation}
-\frac{1}{c}\,\vec{\nabla}\times\left(-c\beta\vec{A}+\vec{\gamma}\,\varphi\right)+\frac{1}{c}\frac{\partial}{\partial t}(\vec{\gamma}\times\vec{A})=-\frac{c}{\hbar}\,\vec{m}\times\vec{A}\,,
\end{equation}
and
\begin{equation}
-\vec{\nabla}\times(\vec{\gamma}\times\vec{A})-\frac{1}{c^2}\frac{\partial}{\partial t}\left(-c\beta\vec{A}+\vec{\gamma}\varphi\right)-\vec{\nabla}\left(\beta\frac{\varphi}{c}-\vec{\gamma}\cdot\vec{A}\right)=-\frac{m_\ell c}{\hbar}\,\vec{A}+\frac{\vec{m}_t}{\hbar}\varphi\,.
\end{equation}
If we define the electric and magnetic fields as
\begin{equation}
\vec{E}=\mu(-c\beta\vec{A}+\vec{\gamma}\,\varphi)\,,\qquad\qquad  \vec{B}=-\mu\vec{\gamma}\times\vec{A}\,,\qquad\qquad \Lambda=\mu\left(\beta\frac{\varphi}{c}-\vec{\gamma}\cdot\vec{A}\right)\,,
\end{equation}
then eqs.(61) - (64), read
\begin{equation}
\vec{\nabla}\cdot\vec{E}=-\frac{\partial\Lambda}{\partial t}-\mu^2\varphi+\frac{c^2\mu}{\hbar}\,\,\vec{m}_t\cdot\vec{A}\,,
\end{equation}
\begin{equation}
\vec{\nabla}\cdot\,\vec{B}=0\,,
\end{equation}
\begin{equation}
\vec{\nabla}\times\vec{E}=-\frac{\partial\vec{B}}{\partial t}+\frac{c^2\mu}{\hbar}\,\,\vec{m}_t\times\vec{A}\,,
\end{equation}
and
\begin{equation}
\vec{\nabla}\times\vec{B}=\frac{1}{c^2}\frac{\partial\vec{E}}{\partial t}+\vec{\nabla}\Lambda-\mu^2\vec{A}+\mu\,\frac{\vec{m}_t}{\hbar}\,\varphi\,.
\end{equation}
The electric and magnetic fields in eq.(65) are linear functions of $\vec{A}$ and $\varphi$ are new to electrodynamics. They, therefore, represent a new option for electrodynamics.   They can be compared with eqs.(42) - (45). Since we use the Dirac matrices in these definitions, we suggest calling these fields the Dirac-Maxwell electromagnetic fields. Interestingly, eqs.(66) - (69) reveal the interaction of the vector mass with the electromagnetic potentials. They represent a  generalization of the Maxwell-Proca equations for vector mass and $\Lambda\ne 0$\,. Recall that only a charged particle has the opportunity to interact with the electromagnetic field, however, a massive photon can interact with its vector mass somehow. The presence of such an interaction could help explain how the neutron, inside a nucleus, interacts with the neighboring protons.

The electric, magnetic fields and scalar function ($\Lambda$) in eq.(65) are invariant under the gauge transformations provided
\begin{equation}
c\beta\vec{\nabla}f+\vec{\gamma}\frac{\partial f}{\partial t}=0\,,\qquad \vec{\gamma}\times\vec{\nabla}f=0\,,\qquad \beta\frac{\partial f}{\partial t}+c\vec{\gamma}\cdot\vec{\nabla}f=0\,.
\end{equation}
The above equations are satisfied if $f$ satisfies the wave equation
\begin{equation}
\nabla^2f-\frac{1}{c^2}\frac{\partial^2 f}{\partial t^2}=0\,.
\end{equation}
Therefore, eqs.(66) - (69) are gauge invariant too with the conditions
\begin{equation}
cm_\ell\vec{\nabla}f+\vec{m}_t\frac{\partial f}{\partial t}=0\,,\qquad m_\ell\frac{\partial f}{\partial t}+c\vec{m}_t\cdot\vec{\nabla}f=0\,,\qquad \vec{m}_t\times\vec{\nabla}f=0\,.
\end{equation}
Equations (70) and (72) imply  that
\begin{equation}
m_\ell\vec{\gamma}=\vec{m}_t\beta\,,
\end{equation}
which due to Dirac velocity prescription, $\vec{v}=\beta\vec{\gamma}$, yields
\begin{equation}
m_\ell\vec{v}=\vec{m}_tc\,.
\end{equation}
It agrees once again with eq.(41b).

\section{Concluding remarks}
Different scenarios in which the photon mass is quaternionic are investigated. The massive electrodynamics with quaternionic mass yields  quantized and generalized axion electrodynamics. The use of the quaternionic mass in massive electrodynamics is found to be equivalent to introducing  interaction with the electromagnetic field. Interestingly,  the energy conservation equation of the massive electrodynamics is not influenced by the presence of the two masses. A system of equations describing the interaction of the axion field with the quantum particle is provided.

The interaction of the quantum particle with the electromagnetic field can be obtained from that of a free particle with quaternionic mass provided one defines $c\,\vec{m}_t\equiv q\vec{A}$ and $m_\ell c^2\equiv q\varphi$. A new equation supporting the de Broglie hypothesis is derived where a new concept of vector mass for a wave is introduced. Therefore, the mass besides being the source of the matter-wave could also be a source of the electromagnetic field. Interestingly, the quaternionic Dirac equation with quaternionic mass yields the Proca-Maxwell equations. New definitions for the electric and magnetic fields are proposed. The presence of a quaternionic mass in Maxwell's theory is shown to give the same effects as that attributed to axions. It is shown that the new massive electrodynamics yields Maxwell-Proca equations that are also shown to be equivalent to axion electrodynamics.

\end{document}